\documentclass[fleqn,10pt]{wlscirep}
\usepackage[utf8]{inputenc}
\usepackage[T1]{fontenc}
\usepackage{amsmath}
\usepackage{lineno}

\usepackage{todonotes}
\usepackage{xcolor} 
\usepackage{xspace}

\newcommand{\hlsfml}{\texttt{hls4ml}\xspace}

\newcommand{\pt}{\ensuremath{p_{\mathrm{T}}}\xspace}
\newcommand{\GeV}{\ensuremath{\,\text{Ge\hspace{-.08em}V}}\xspace}
\newcommand{\TeV}{\ensuremath{\,\text{Te\hspace{-.08em}V}}\xspace}

\title{LHC physics dataset for unsupervised New Physics detection at 40 MHz}

\author[1]{Ekaterina Govorkova}
\author[1,*]{Ema Puljak}
\author[1]{Thea Aarrestad}
\author[1]{Maurizio Pierini}
\author[1,$\ddagger$]{Kinga Anna Wo\'zniak}
\author[2,$\dag$]{Jennifer Ngadiuba}

\affil[1]{European Organization for Nuclear Research (CERN), CH-1211 Geneva 23, Switzerland}
\affil[2]{Fermi National Accelerator Laboratory, Batavia, IL 60510, USA}

\affil[*]{Corresponding author: Ema Puljak (ema.puljak@cern.ch)}

\affil[$\dag$]{Also at California Institute of Technology, USA}
\affil[$\ddagger$]{Also at University of Vienna, Austria}

\begin{abstract}
In particle detectors at the Large Hadron Collider, tens of terabytes of data are produced every second from proton-proton collisions occurring at a rate of 40 megahertz. This data rate is reduced to a sustainable level by a real-time event filter processing system which decides whether each collision event should be kept for further analysis or be discarded. We introduce a dataset of proton collision events that emulates a typical data stream collected by such a real-time processing system, pre-filtered by requiring the presence of at least one electron or muon. This dataset could be used to develop novel event selection strategies and assess their sensitivity to new phenomena. In particular, by publishing this dataset we intend to stimulate a community-based effort towards the design of novel algorithms for performing unsupervised New Physics detection, customized to fit the bandwidth, latency and computational resource constraints of the real-time event selection system of a typical particle detector.

\end{abstract}
\begin{document}

\flushbottom
\maketitle

\thispagestyle{empty}

\section*{Background \& Summary}
\begin{figure}[!ht]
    \centering
    \includegraphics[width=0.99\textwidth]{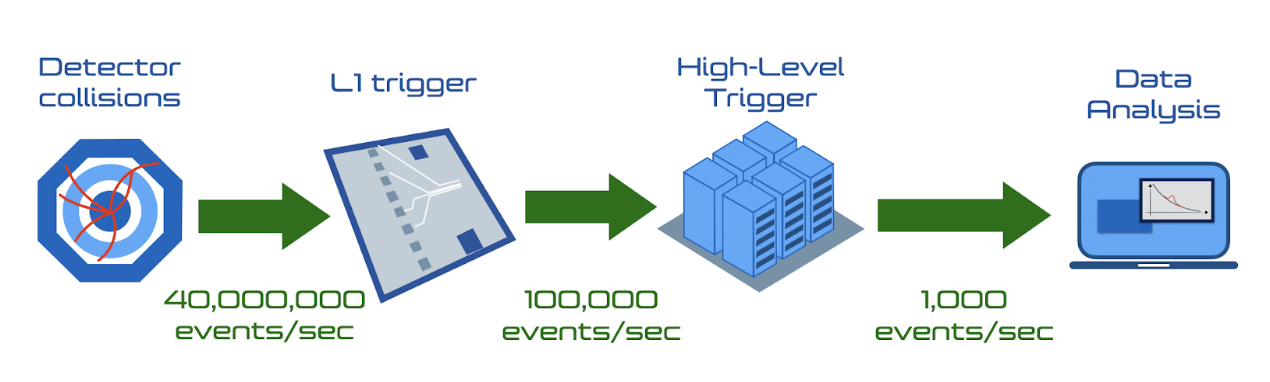}
    \caption{The real-time data processing flow of the ATLAS and CMS experiments: 40M collisions are produced every second and processed by the hardware-based event selection system, consisting of algorithms implemented as logic circuits on custom electronic boards. Of these events, 100K are accepted and passed to the second selection stage, the HLT, which selects about 1000/sec for offline physics studies.\label{fig:data_flow}}
\end{figure}

The proton beams of the CERN Large Hadron Collider (LHC) cross paths 40 million times per second in each of the experimental halls, each time generating a {\it collision event}. In each such event, multiple proton pairs collide, generating thousands of particles in the detectors located at the center of each hall. Detector sensors record the flow of emerging particles in the form of electronic signals, which globally amount to ${\cal O}(1~\mathrm{MB})$ of information. The resulting data throughput of about 40 TB/sec is too large to be recorded. This is why a typical detector, e.g., the two detectors (ATLAS~\cite{ATLAS} and CMS~\cite{CMS}) used to discover the Higgs boson~\cite{CMS_higgs,ATLAS_higgs}, processes data in real-time to select a small fraction of events (about 1000/s), compatible with downstream computing resources. This filtering system, usually referred to as the {\it trigger}, consists of a two-stage selection, as illustrated in Fig.~\ref{fig:data_flow}. The first selection stage, the Level-1 trigger (L1T), runs a set of algorithms deployed as logic circuits on custom electronic boards equipped with field-programmable gate arrays (FPGAs). This stage rejects more than 98\% of the events, reducing the incoming data stream to 100K events/s. Due to the short time interval between two collisions (25 ns) and limited buffer capabilities, the entire pipeline of L1T algorithms has to be executed within ${\cal O}(1)~\mathrm{\mu s}$. The second stage, called the High-Level Trigger (HLT), consists of a computer farm processing events on commercial CPUs, running hundreds of complex selection algorithms within ${\cal O}(100)~\mathrm{ms}$. The trigger selection algorithms are designed to guarantee a high acceptance rate for the physics processes under study.
When designing searches for unobserved physics phenomena, one typically considers specific theory-motivated scenarios. This \textit{supervised} strategy was proven successful when dealing with strongly theoretically motivated searches, e.g., the Higgs boson discovery~\cite{CMS_higgs,ATLAS_higgs}. However, this approach might become a limiting factor in the absence of strong theoretical guidance. The ATLAS and CMS trigger systems could be discarding interesting events, jeopardizing the possibility of new physics discoveries.

Therefore, many recent works investigated {\it unsupervised} and {\it semi-supervised} approaches to data selection and analysis, focusing on Anomaly Detection (AD) strategies based on Deep Learning (DL) algorithms. These studies aim to learn a metric directly from the LHC data, with capability of ranking the events by typicality. One could identify the outliers of this typicality metric distribution as a subset of the collected data potentially enriched with rare (possibly unobserved) physics processes. Extensive reviews of several proposed methods are given in Refs.~\cite{Kasieczka:2021xcg,darkmachines} and references therein.

This effort, mainly targeting offline data analysis, should be paired with a similar effort to integrate AD algorithms in the trigger system of the LHC experiments, possibly already at the L1T. There, it would be possible to present an unbiased dataset to the AD algorithm, before discarding any event~\cite{Cerri:2018anq,Knapp:2020dde}. One could then collect rare event topologies in a special data stream, similar to what was done at CMS with the \textit{exotica hotline}~\cite{symmetry-magazine,Francesco:1306501} during the first year of data taking at the LHC. By studying these events, one could formulate new theoretical models of new physics phenomena, that could be tested in future data-taking campaigns. 

While the focus so far has been on the HLT, this strategy would be more effective if deployed in the L1T, before any selection bias is introduced. Each L1T event has to be processed within a few microseconds and therefore the trigger decision is taken by algorithms hard-coded in the hardware electronics as logic circuits. Algorithm complexity, and possibly accuracy, could be enhanced by deploying DL algorithms in the L1T FPGAs. To do so, the \hlsfml library~\cite{Duarte:2018ite,aarrestad2021fast,AutoQ} was introduced as a tool to translate a given DL model into an electronic circuit. By integrating the full network on the FPGA, \hlsfml privileges inferece speed and is ideal for small networks with ${\cal O}(100~\mathrm{ns})$ latency. 

Given this, all the ingredients are there to make it possible for the community to design an optimal AD strategy for the L1T system. In order to stimulate this effort as a community-based initiative, similarly to what was done with the LHC Olympics~\cite{Kasieczka:2021xcg} and the Dark Machine data challenge~\cite{darkmachines}, we present a new dataset designed to mimic a typical data stream from a L1T system.

\section*{Methods}

\subsection*{Physics content of the dataset}
The proton-proton collisions taking place at the LHC can lead to the production and observation of many different processes predicted by the Standard Model of particle physics~\cite{GLASHOW1961579,PhysRevLett.19.1264,Salam}. A brief summary of the SM particle content can be found in Refs.~\cite{SM-quantamagazine,Quigg:2005wa}. The rate at which each of these processes occur can be calculated within the SM mathematical framework and then validated by the measurements performed by the experiments~\cite{CMS-SMP-summary}. 
In this paper, we focus on events containing electrons ($e$) and muons ($\mu$), light particles that, together with taus ($\tau$) and their neutrino partners, form the three lepton families. In principle, we could have considered a dataset with no filter. While this would certainly be a more realistic representation of an unbiased L1T stream, generating such a dataset requires computing resources beyond our capabilities. Instead, we decided to use the lepton filter and make the dataset simulation tractable. 

Within the limited size of a typical LHC detector, electrons and muons are stable particles, i.e., they do not decay in the detector and they are directly observed while crossing the detector material. Instead, $\tau$ leptons are much heavier than electrons and muons. They quickly decay into other particles. In a fraction of these decays, $e$ and $\mu$ are produced. At the LHC, the most abundant source of high-energy leptons is the production of $W$ and $Z$ bosons~\cite{Zyla:2020zbs} which are among the heaviest SM particles with a mass of $\sim 80$ and $\sim90\GeV$, respectively. Once produced, they quickly decay into other particles, notably leptons. $W$ and $Z$ bosons are mainly produced directly in proton collisions. A sizable fraction of $W$ bosons originate from the decay of top quarks ($t$) and anti- quarks ($\overline{t}$). Being the top quark heavy and highly unstable, it quickly decays into a $W$ boson and a bottom quark, giving rise to signatures with only jets or with one $e$, $\mu$, or $\tau$, a neutrino and a jet. Leptons can originate from more rare $W$ and $Z$ production, such as from the decay of Higgs bosons or multi-boson production. Given the small production probability of these processes, we ignore them in this study.

An important source of leptons comes from the production of light quarks (up, down, charm, strange, and bottom) and gluons, predicted by Quantum Chromodynamics (QCD)~\cite{ellis_stirling_webber_1996}. As these quarks and gluons have a net colour charge and cannot exist freely due to colour-confinement, they are not directly observed. Instead, they come together to form colour-neutral hadrons, a process called hadronisation that leads to a collimated spray of hadrons called a {\it jet}. A jet is usually defined by the algorithm used to cluster such spray of particles, e.g. the anti-$k_T$ algorithm~\cite{Cacciari:2008gp}. Rarely, leptons can be produced inside jets, typically from the decay of unstable hadrons. On the other hand, QCD multijet production is by far the most abundant process at the LHC, thereby this contribution is sizable and taken into account.

\subsection*{Dataset}
\label{sec:data}

The processes listed above are the main contributors to an $e$ or $\mu$ data stream, i.e., the set of collision events selected for including an $e$ or $\mu$ with energy above a defined threshold. One of the datasets presented in this paper consist of the simulation of such a stream. In addition, benchmark examples of new lepton-production processes are given. These processes consist of the production of postulated, but still unobserved particles. They serve as examples of data anomalies that could be used to validate the performance of an AD algorithm. Details on these processes can be found in Refs.~\cite{Cerri:2018anq,Knapp:2020dde}.

Moreover, we published a {\it blackbox} dataset containing a mixture of SM processes and a {\it secret} signal process. Events in this dataset are uniquely labeled by an event number, which allows us to link each event to the corresponding {\it ground-truth} dataset, containing a set of 0 (for SM events) and 1 (for new physics events) bits. The ground truth dataset is stored on a private cloud storage area at CERN. Not publishing this dataset, we can assure that the distributed {\it blackbox} is unlabeled for external developers. It is intended to be used to independently validate the performance of AD algorithms developed based on this dataset. We consider the following Standard Model (SM) processes, with their relative contribution to the dataset listed in parenthesis:
\begin{itemize}
    \item Inclusive $W$ boson production, where the $W$ boson decays to a charged lepton ($\ell$) and a neutrino ($\nu)$, (59.2\% of the dataset). The lepton could be a $e$, $\mu$, or $\tau$ lepton. 
    \item Inclusive $Z$ boson production, with $Z \to \ell \ell \,\,(\ell = e, \mu, \tau )$ (6.7\% of the dataset),
    \item $t\bar{t}$ production (0.3\% of the dataset), and
    \item QCD multijet production (33.8\% of the dataset).
\end{itemize}
These four samples are mixed to form a realistic data stream
populated by known SM processes (collectively referred to as {\it background}), and is provided in Ref.~\cite{zenodo_bkg_training}. An AD algorithm can thus be trained on this sample to learn the underlying structure of the background to identify a new physics signature (the {\it signal}) as an outlier in the distribution of the learned metrics.

To study the performance of AD algorithms, four signal datasets are provided:
\begin{itemize}
    \item A leptoquark (LQ) with mass 80 GeV, decaying to a $b$ quark and a $\tau$ lepton~\cite{zenodo_LQ},
    \item A neutral scalar boson ($A$) with mass 50 GeV, decaying to two off-shell $Z$ bosons, each forced to decay to two leptons: $A \to 4\ell$~\cite{zenodo_AZZ},
    \item A scalar boson with mass 60~GeV, decaying to two tau leptons: $h^0\to \tau \tau$~\cite{zenodo_htautau},
    \item  A charged scalar boson with mass 60~GeV, decaying to a tau lepton and a neutrino: $h^\pm \to \tau \nu$~\cite{zenodo_htaunu}.
\end{itemize}
These samples are generated using the same code and workflow as for the SM events.

\begin{figure}[!ht]
    \centering
    \includegraphics[width=0.60\textwidth]{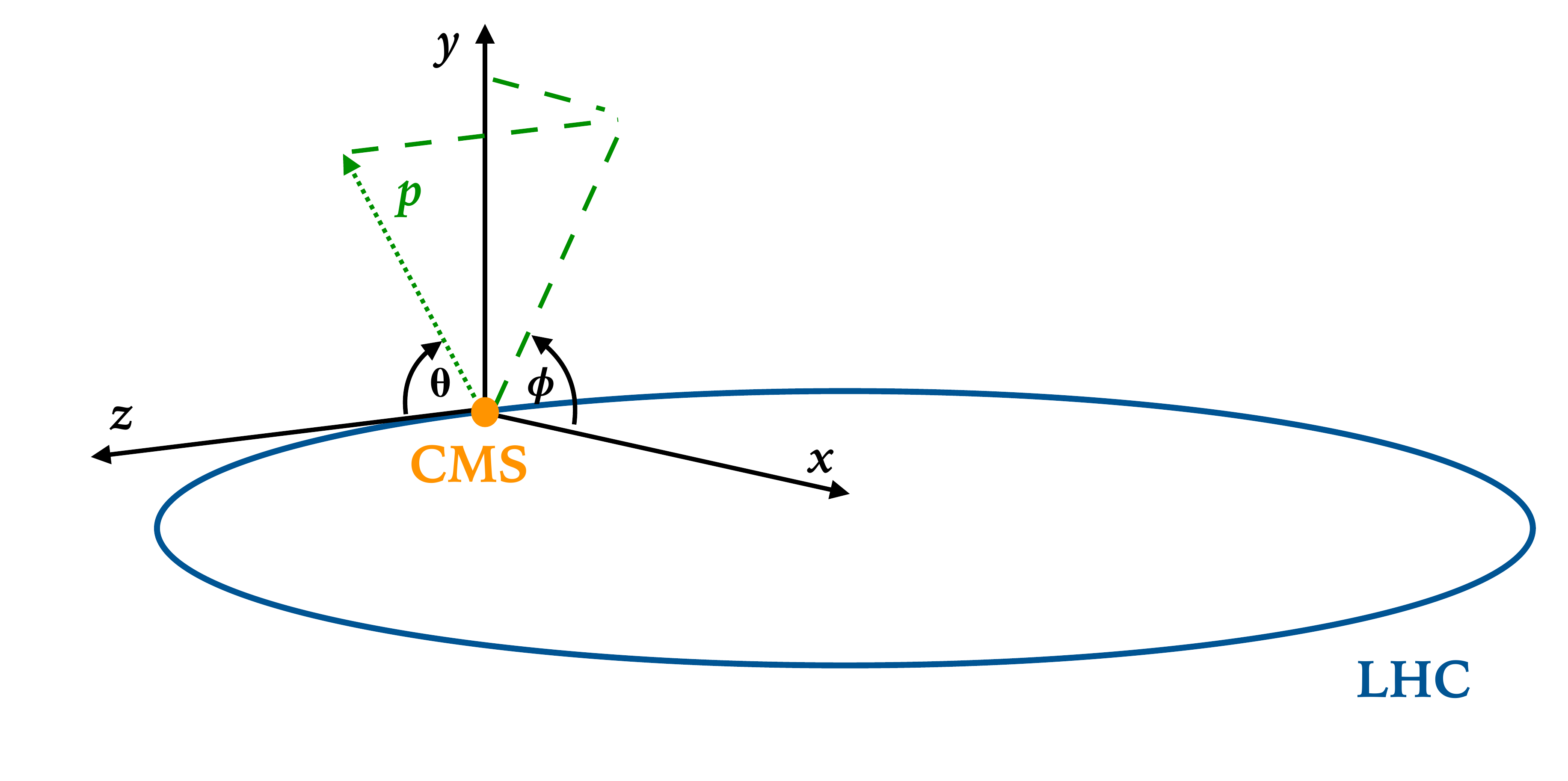}
    \caption{The reference system used to describe the momentum coordinates of the particles in the dataset.\label{fig:ref_system}}
\end{figure}

We use a Cartesian coordinate system with the $z$ axis oriented along the beam axis, the $x$ axis on the horizontal plane, and the $y$ axis oriented upward as shown in Fig.~\ref{fig:ref_system}. The $x$ and $y$ axes define the transverse plane, while the $z$ axis identifies the longitudinal direction. Azimuth angle $\phi$ is computed with respect to the $x$ axis. Its value is given in radiants, in the [$-\pi, \pi$] range. The polar angle $\theta$ is used to compute the pseudorapidity $\eta = -\log(\tan(\theta/2))$. The transverse momentum ($p_T$) is the projection of the particle momentum on the ($x$, $y$) plane. We fix units such that $c=h/2\pi=1$, where $c$ is the speed of light and $h$ is the Planck constant. Charge conjugation is implicit. 

Each event is represented by a list of four-momenta for high-level reconstructed objects: muons, electrons, and jets. In order to emulate the limited bandwidth of a typical L1T system, we consider only the first 4 muons, 4 electrons, and 10 jets in the event, selected after ordering the candidates by decreasing $p_T$. If an event contains less particles, the event is zero-padded to preserve the size of the input, as it is done in realistic L1T systems. Each particle is represented by its $p_{T}$, $\eta$, and $\phi$ values. In addition, we consider the absolute value and $\phi$ coordinates of the missing transverse energy, defined as the vector equal and opposite to the vectorial sum of the transverse momenta of all the reconstructed particles in the event. 

Once generated, events are filtered using a custom selection algorithm, coded in {\tt Python}. We require a reconstructed electron or a muon with $p_T>23$~GeV within $|\eta|<3$ and $|\eta|<2.1$, respectively. Up to ten jets with $p_T>15$~GeV within $|\eta|<4$ are include in the event, together with up to four muons with $|\eta|<2.1$ and $p_T>3$~GeV, up to four electrons with $|\eta|<3$ and $p_T>3$~GeV, and the missing transverse energy (MET) defined as the vector equal and opposite to the vectorial sum of the momenta of all particles reconstructed in the event. Given these requirements, the four SM processes listed above provide a realistic approximation of a L1T data stream.

In total, the SM cocktail dataset consists of 8,209,492 events, of which 4 millions are used to define the training dataset~\cite{zenodo_bkg_training}. The rest is mixed to events from the secret new physics process, to generate the  {\it blackbox} dataset~\cite{zenodo_blackbox}. Together with the particle momenta, this dataset includes the event numbers needed to match each event to its ground-truth bit.

\section*{Data Records}
The publication consists of six data records: one record containing the mixture of SM processes, four separate records for each of the BSM processes listed above and one record containing the  {\it blackbox} data. These are listed in Table~\ref{tab:samples}, together with the total number of events and whether the record is considered to be background, signal or a mixture of the two.

The datasets are revised versions of those utilized in Refs.~\cite{Cerri:2018anq,Knapp:2020dde} and already published on Zenodo~\cite{cerri_olmo_2020_3675199,cerri_olmo_2020_3675203,cerri_olmo_2020_3675206,cerri_olmo_2020_3675210,cerri_olmo_2020_3675196,cerri_olmo_2020_3675159,cerri_olmo_2020_3675190,cerri_olmo_2020_3675178}. They differ from each other on the data format, and the inclusion of a  {\it blackbox} dataset containing secret new physics events. The benchmark new physics datasets have the same physics content as the original datasets, but the included events were generated specifically for this paper. The data records are published on Zenodo~\cite{zenodo_bkg_training,zenodo_AZZ,zenodo_LQ, zenodo_htautau, zenodo_htaunu,zenodo_blackbox}.

\begin{table}[ht]
\centering
\begin{tabular}{|l|l|l|}
\hline
Sample name & Number of samples & Type \\
\hline
SM processes~\cite{zenodo_bkg_training}& 4,000,000 & B \\
\hline
$LQ \to b\tau$~\cite{zenodo_LQ} & 340,544 & S \\
$A \to 4\ell$~\cite{zenodo_AZZ} &55,969 & S \\
$h^0\to \tau \tau$~\cite{zenodo_htautau} & 691,283 & S \\
$h^\pm \to \tau \nu$~\cite{zenodo_htaunu} & 760,272 & S \\
\hline
 {\it blackbox}~\cite{zenodo_blackbox}& 4,210,492 & S+B \\
\hline
\end{tabular}
\caption{\label{tab:samples} The names and corresponding Zenodo reference for each data set, the total number of collision events and the dataset type (S for signal and B for background).}
\end{table}

The data records are provided in Hierarchical Data Format version 5 (HDF5), and contains 3 datasets: "Particles", "Particles\_Classes", "Particles\_Names". The "Particles" dataset has a shape (N, 19, 4), where N is the number of events listed for each sample in Table~\ref{tab:samples}. The second index runs over the different physics objects in the events: MET, 4 electrons, 4 muons, 10 jets. Its cardinality (19) is the maximum number of objects per event. If less objects are present, the event is zero padded in such a way that the 1st, 5th, and 9th positions correspond to the highest-$p_T$ electron, muon, and jet, respectively. The last index (with cardinality 4) runs over the three features describing each physics object and a particle type index, which is equal to 1, 2, 3 and 4 for MET, electron, muon and jet correspondingly. Zero padding is done inclusively, e.g. for zero-padded particles particle type index is set to zero. The features are ordered as described in the "Particles\_Names" dataset: $\pt, \eta, \phi$. The  {\it blackbox} sample includes an additional dataset ("EvtId") with dimension~(N), containing an event ID which allows us to match each event to its ground truth (signal or background).

\section*{Technical Validation}

\begin{figure}[!ht]
    \centering
    \includegraphics[width=0.3\textwidth]{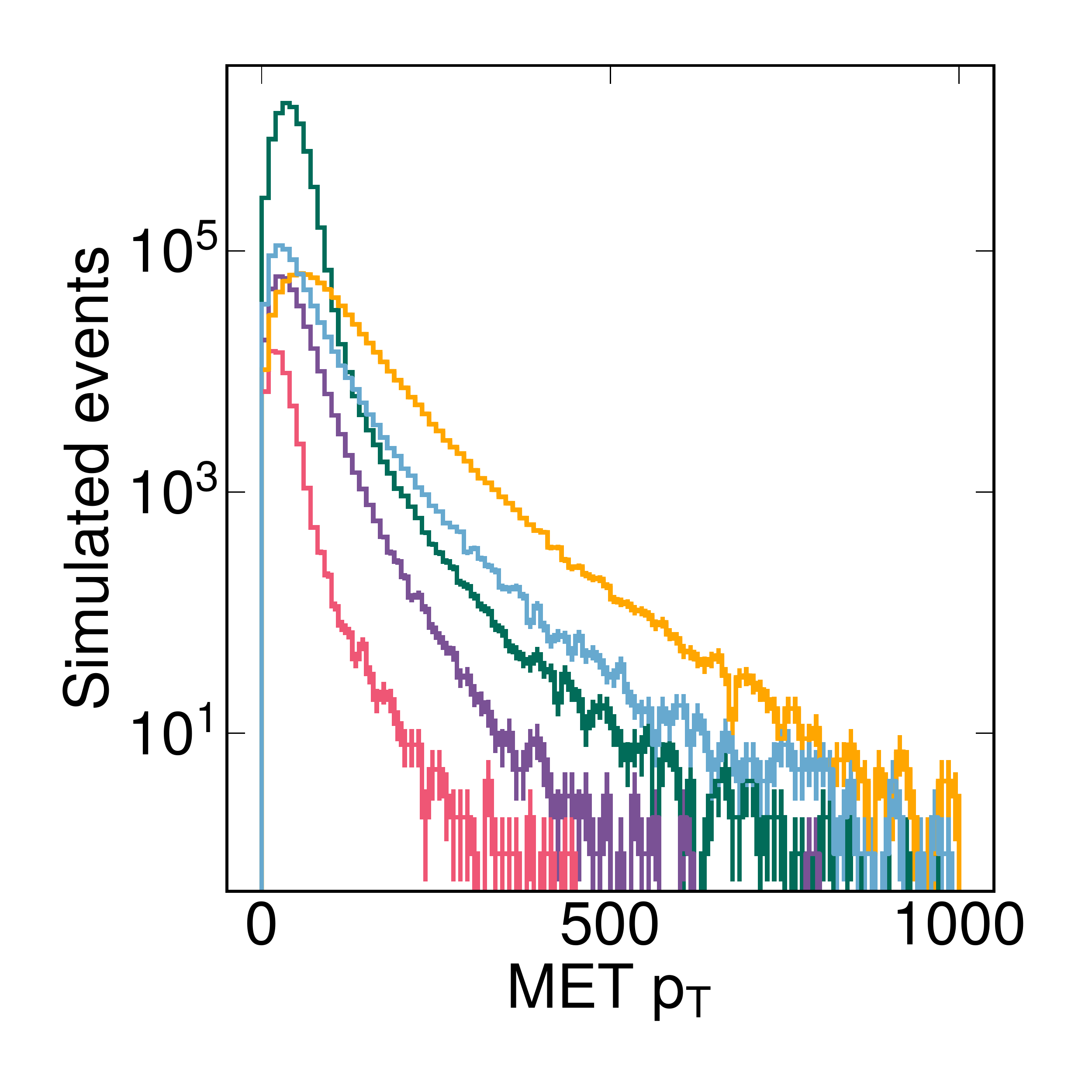}
    \includegraphics[width=0.3\textwidth]{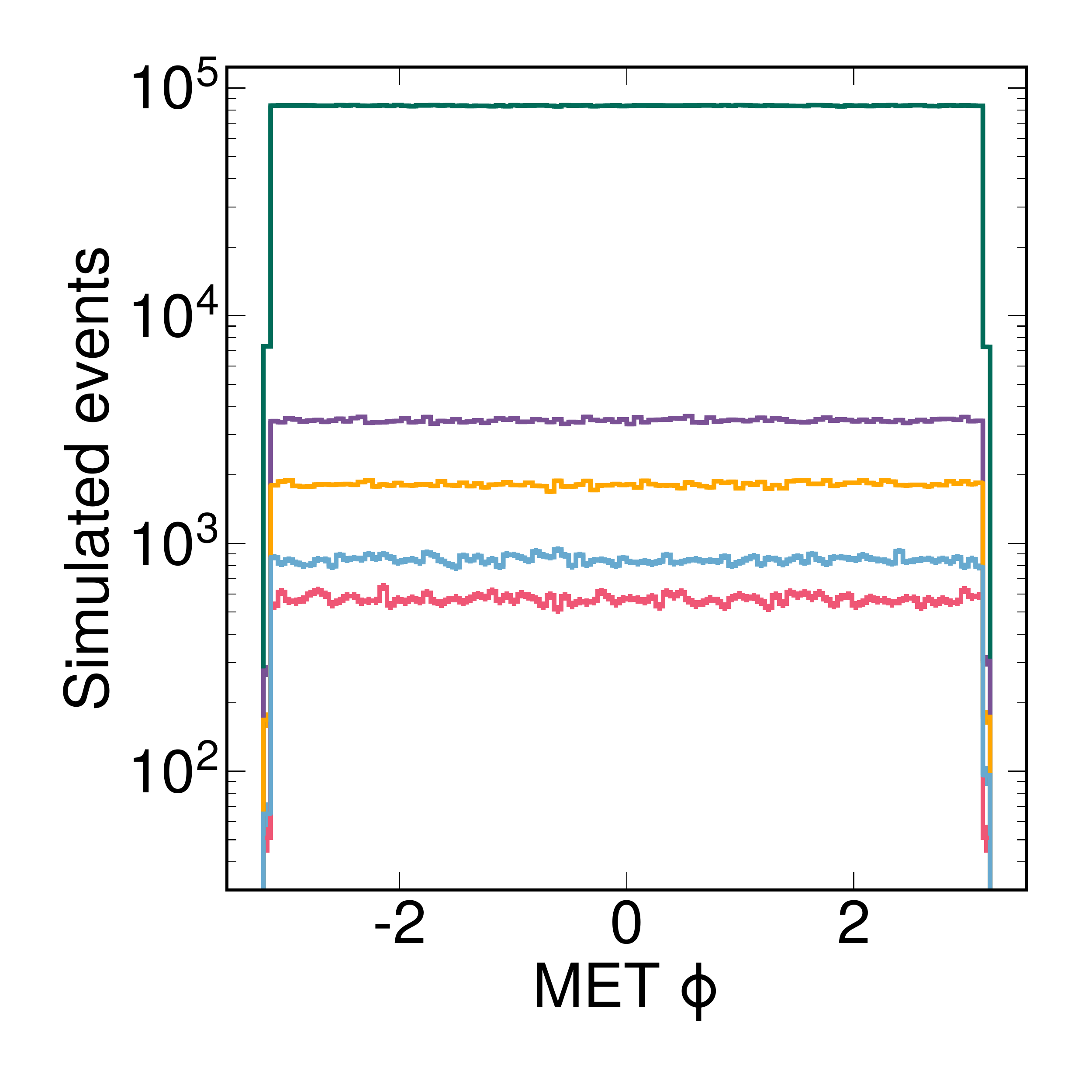}
    \includegraphics[width=0.3\textwidth]{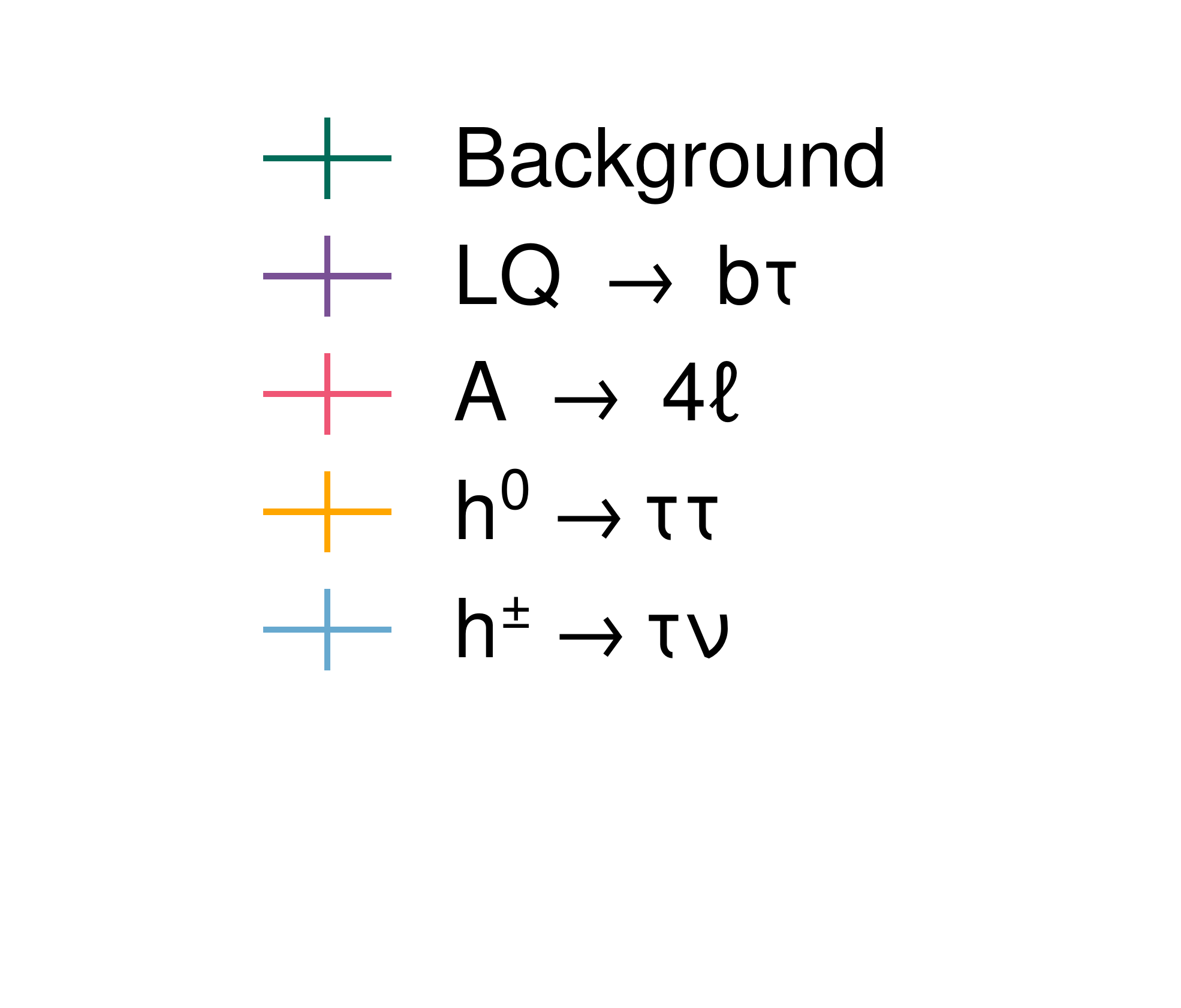}

    \includegraphics[width=0.3\textwidth]{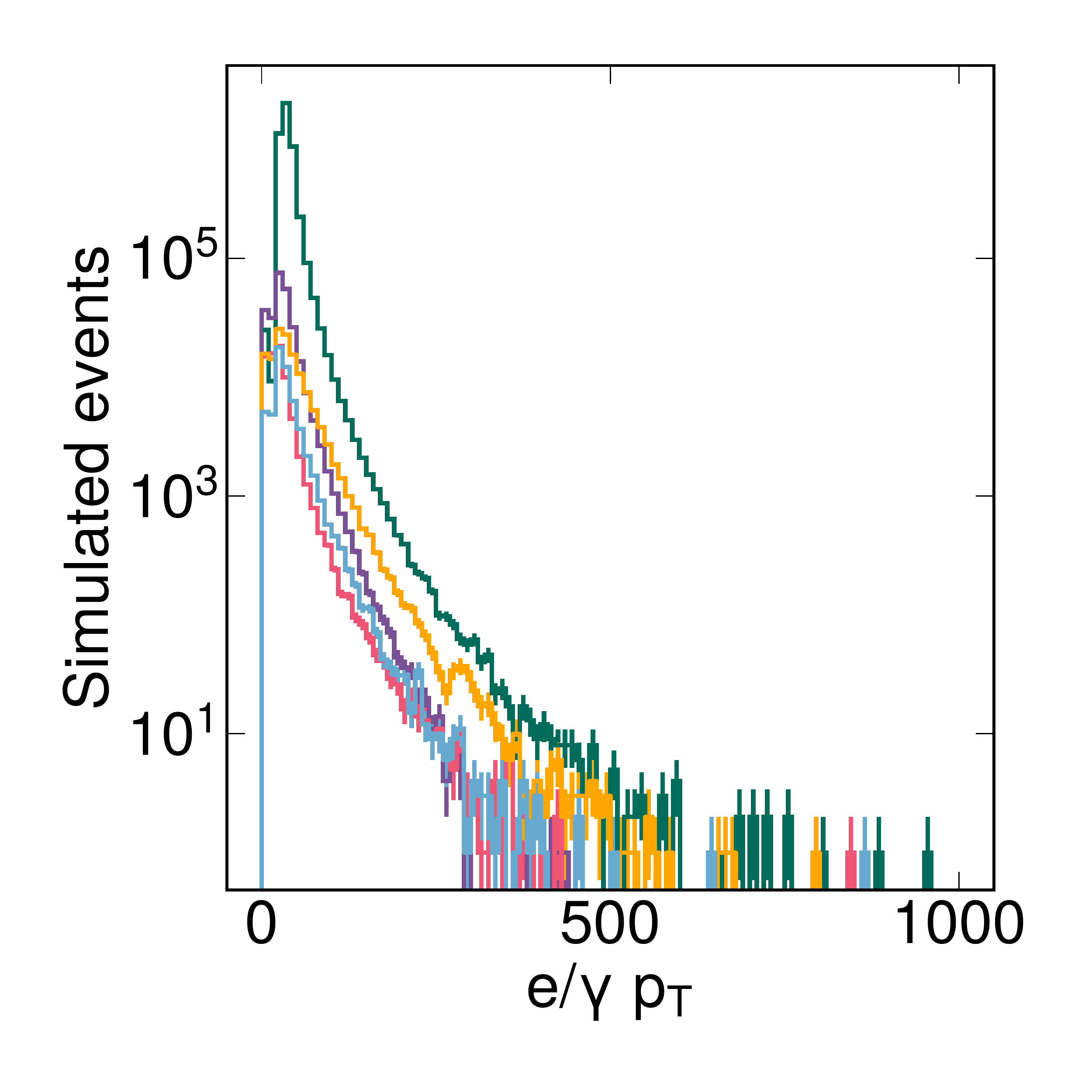}
    \includegraphics[width=0.3\textwidth]{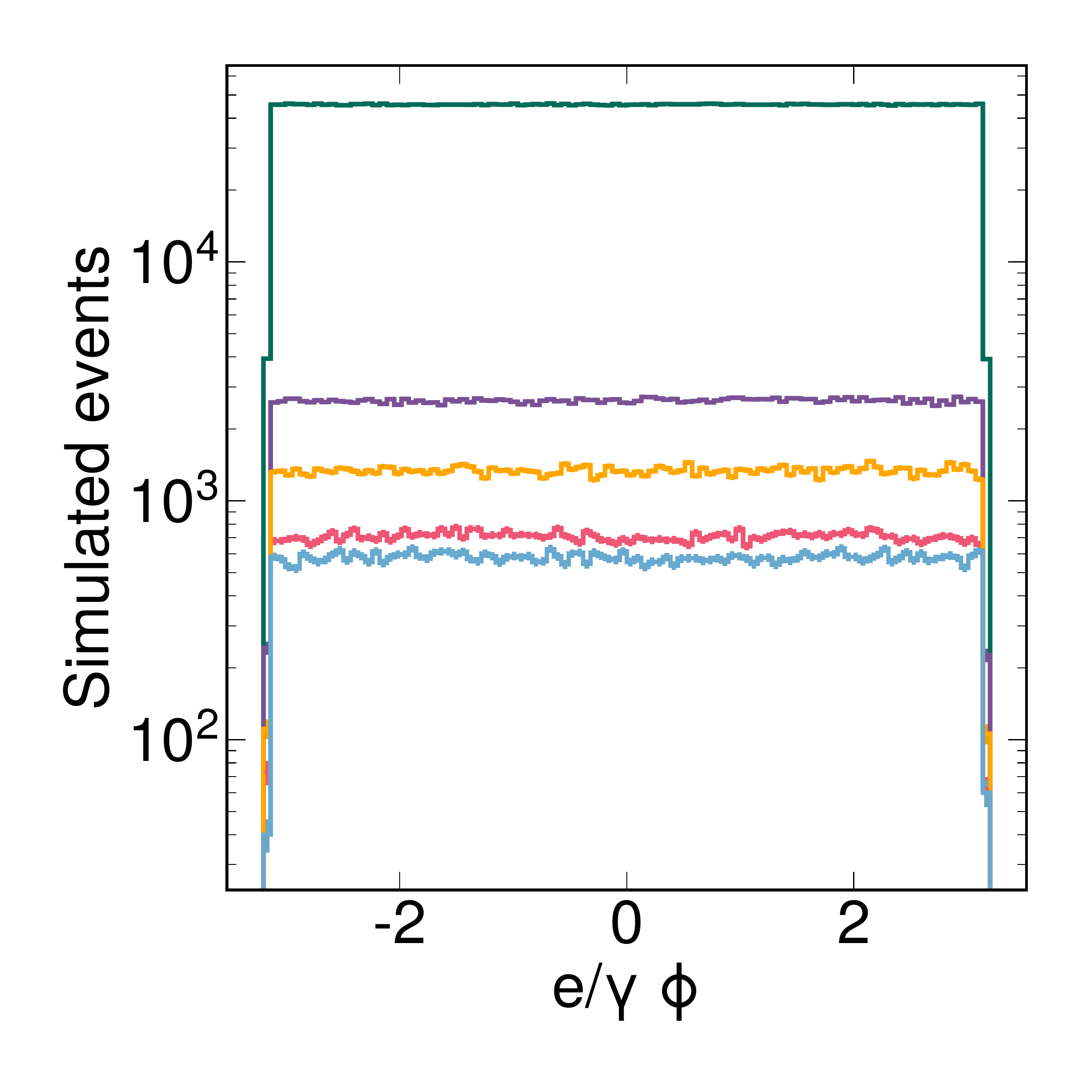}
    \includegraphics[width=0.3\textwidth]{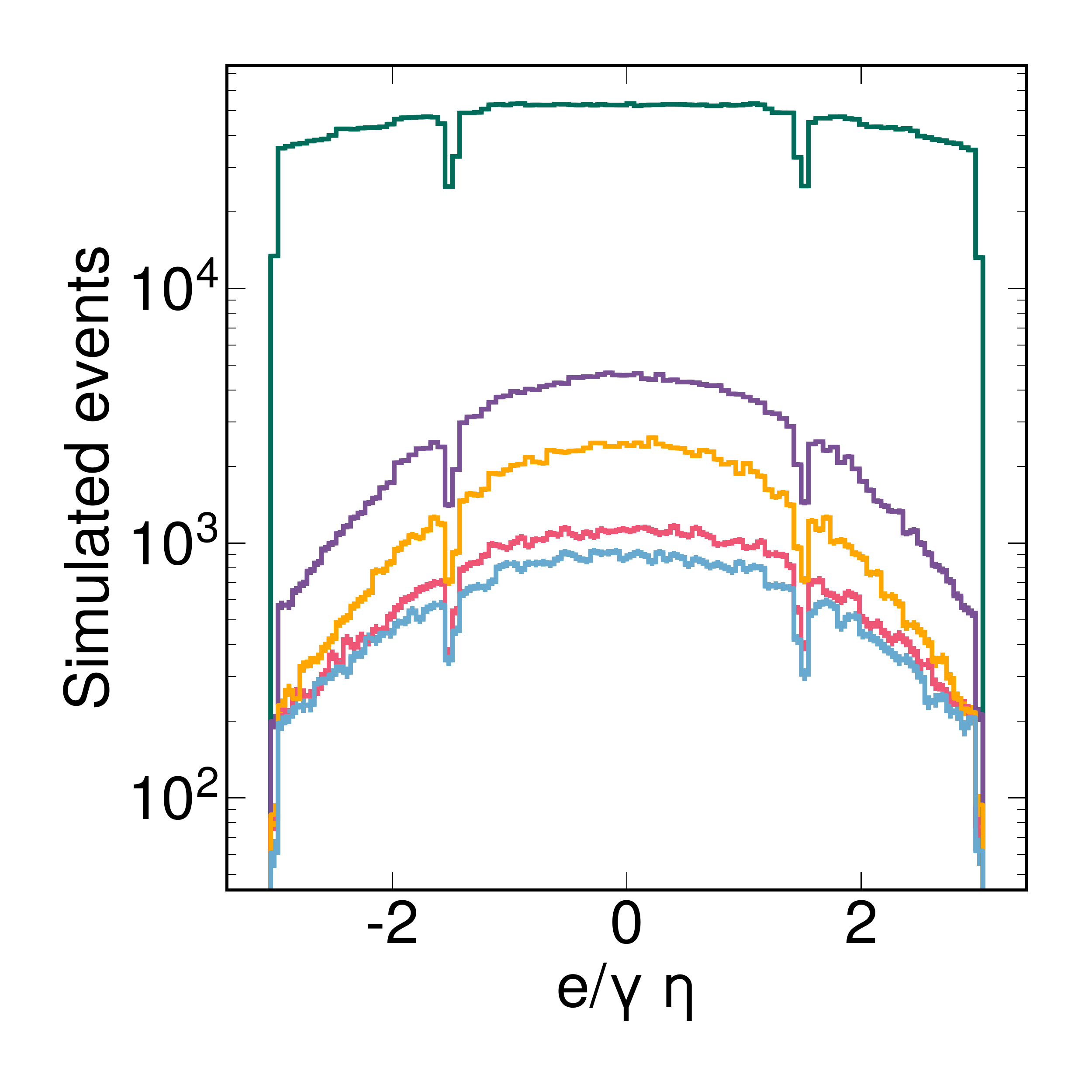}

    \includegraphics[width=0.3\textwidth]{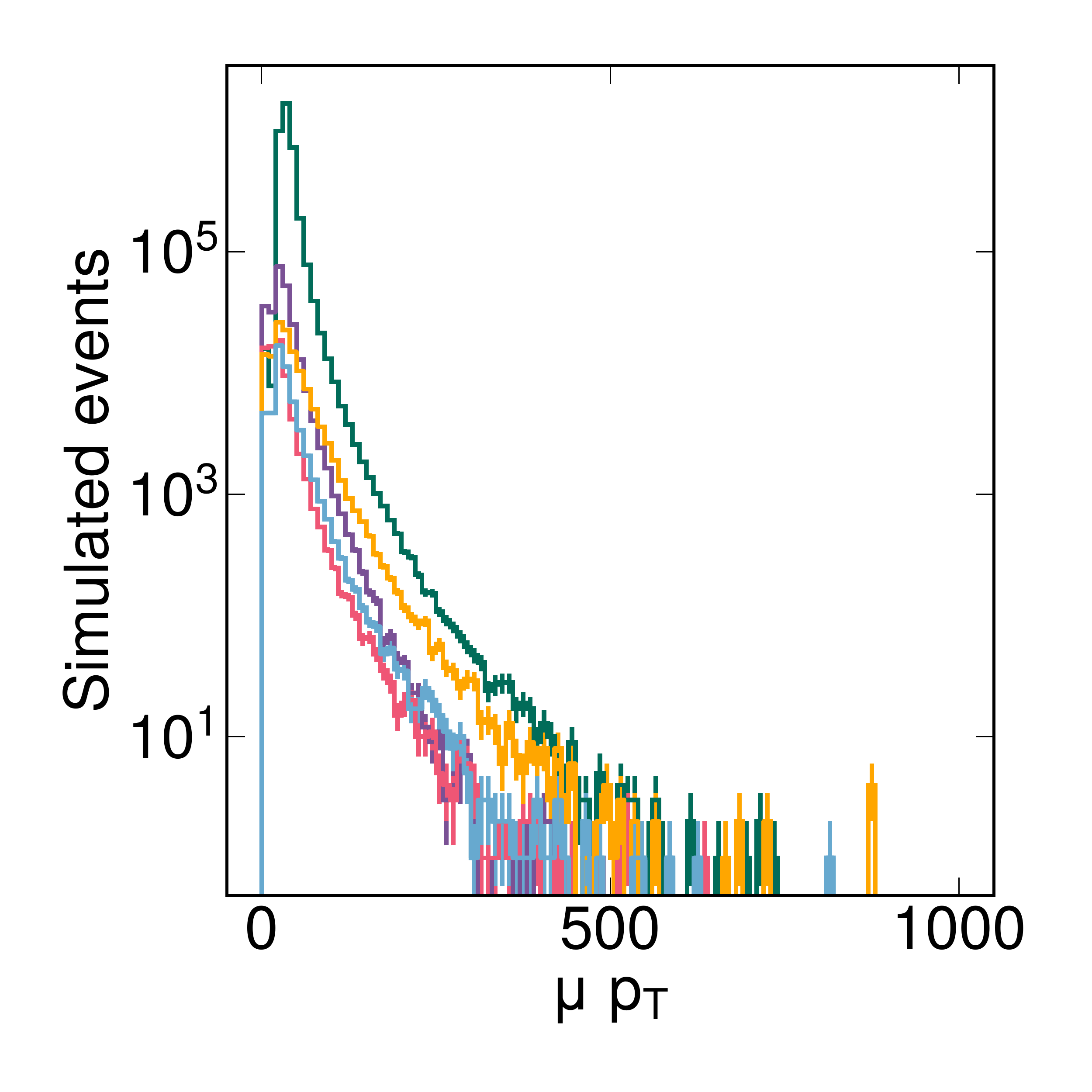}
    \includegraphics[width=0.3\textwidth]{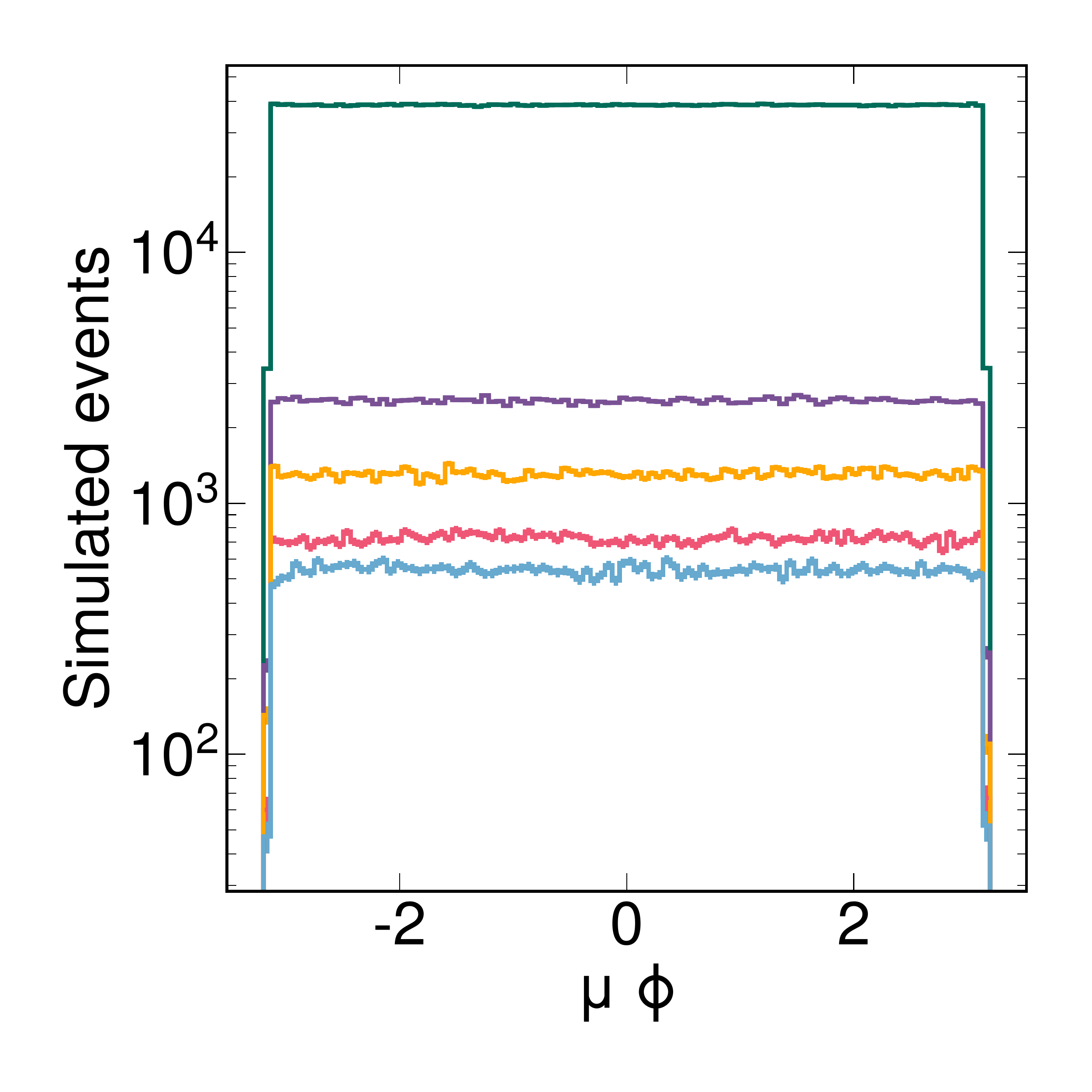}
    \includegraphics[width=0.3\textwidth]{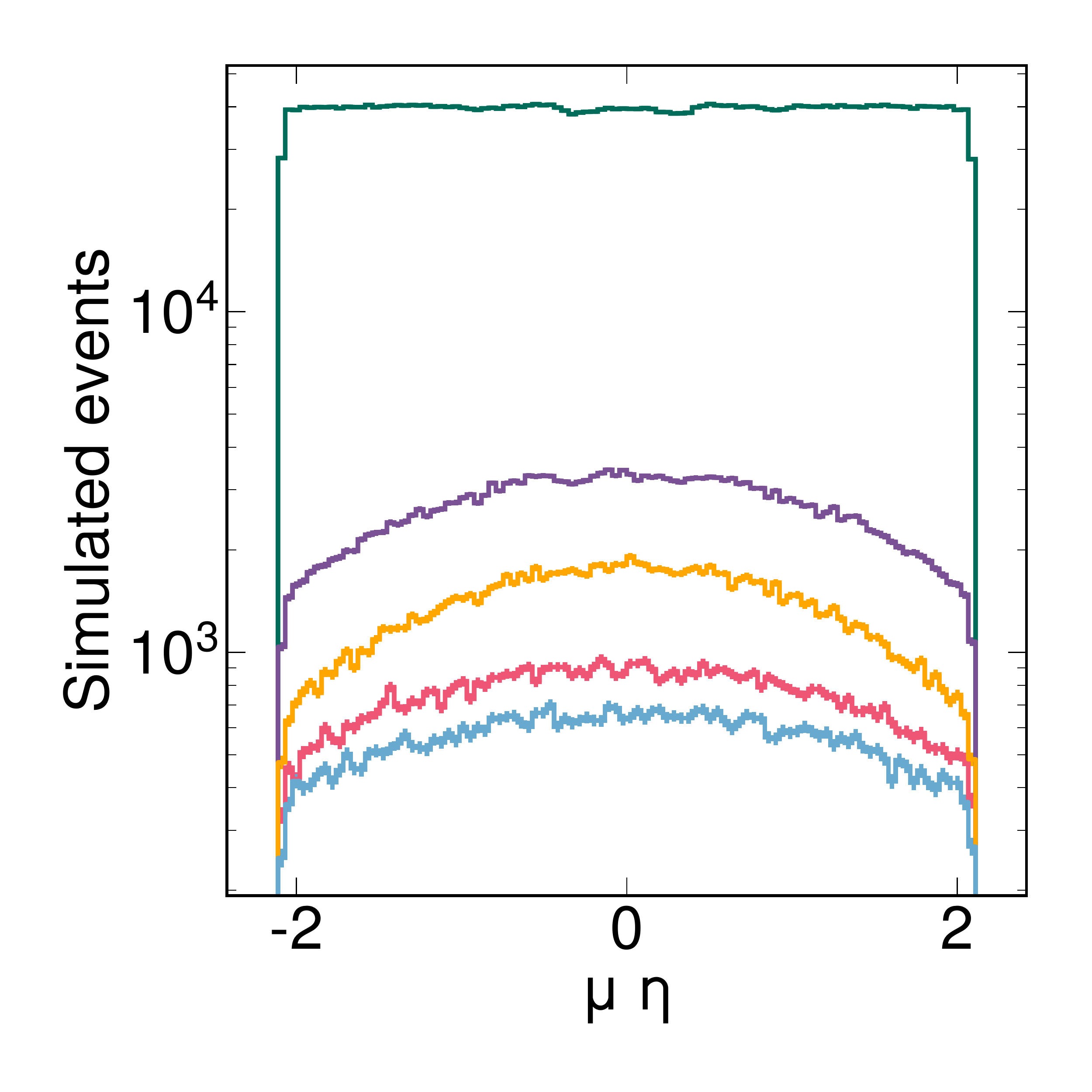}

    \includegraphics[width=0.3\textwidth]{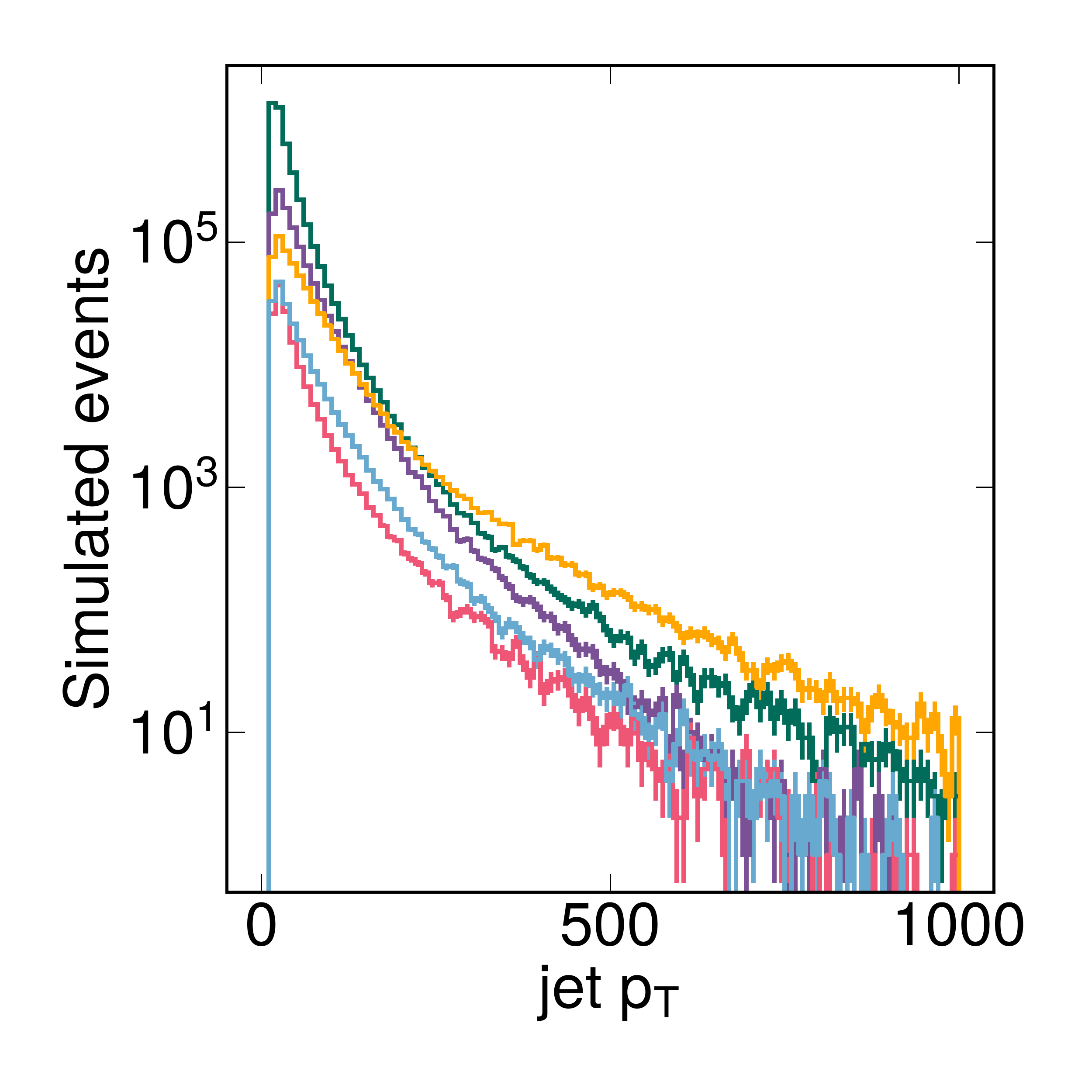}
    \includegraphics[width=0.3\textwidth]{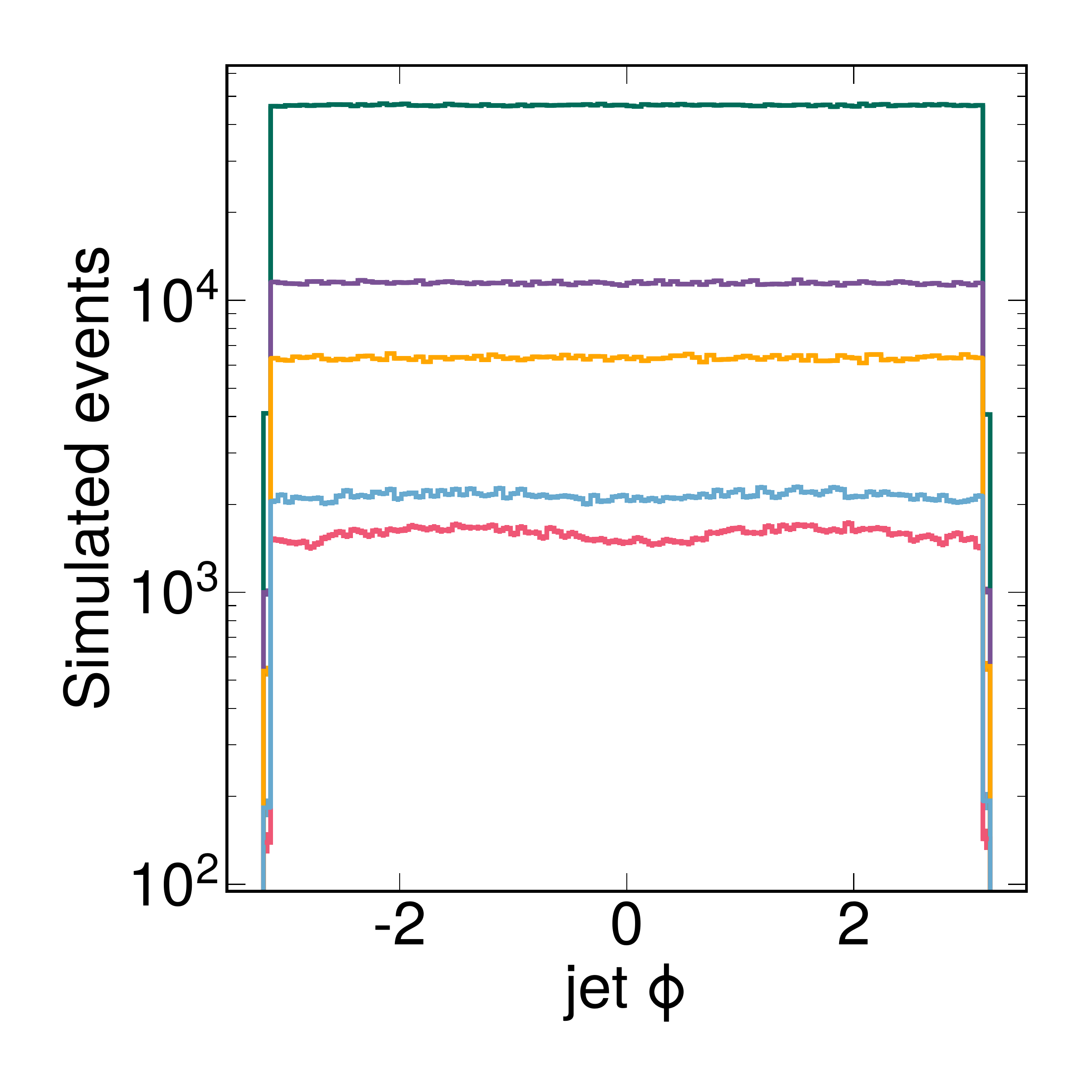}
    \includegraphics[width=0.3\textwidth]{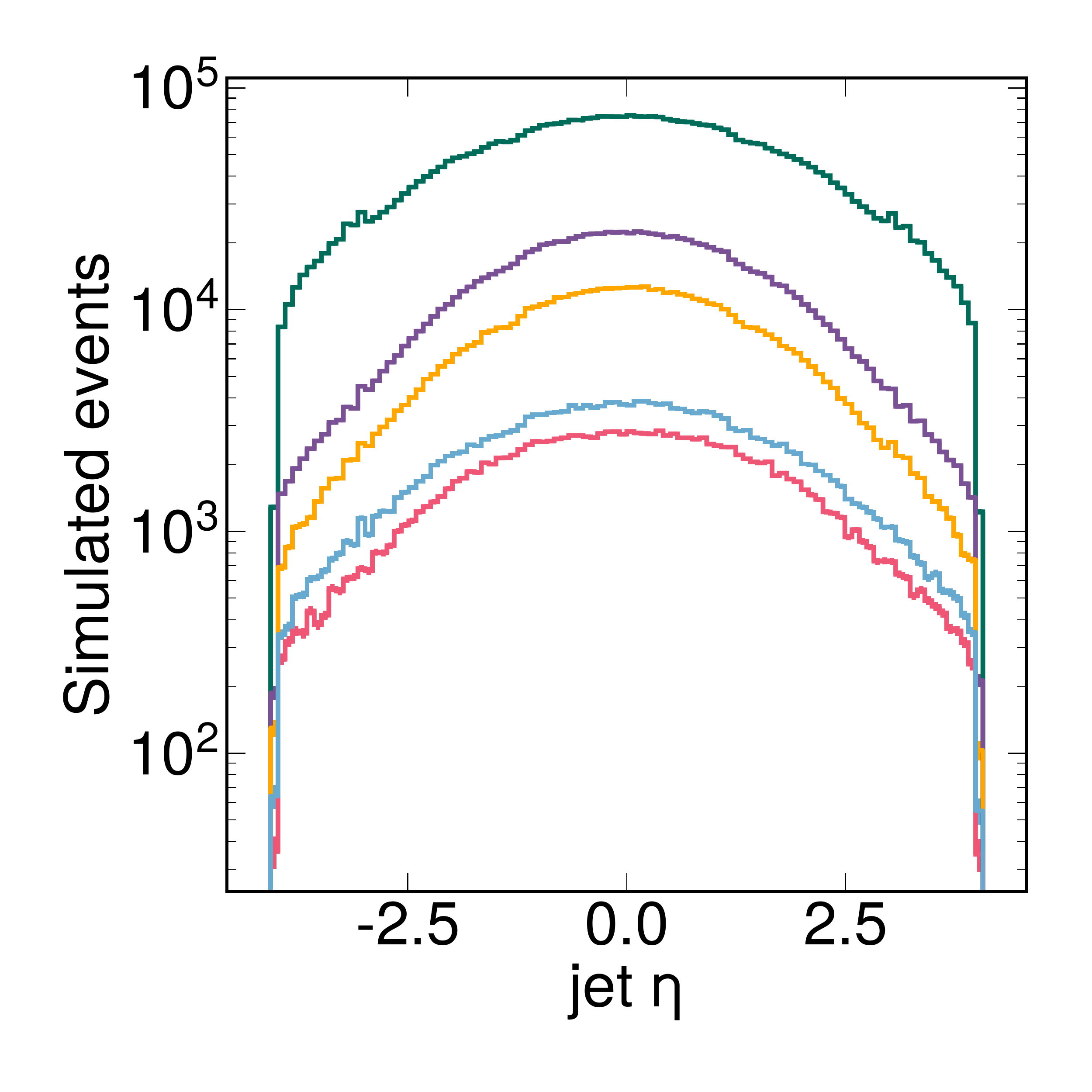}

    \caption{Distribution of the $p_T$ (left), $\phi$ (center) and $\eta$ (right) coordinates of the physics objects entering the dataset, for  missing transverse energy, MET (top row), electrons (second row), muons (third row) and jets (bottom row).\label{fig:particle_features}}
\end{figure}

The distributions of the features for the SM processes and for the chosen BSM models are shown in Fig.~\ref{fig:particle_features}. All expected features are observed, e.g., the detector $\phi$ symmetry, the detection inefficiency in $\eta$ in the transition regions between detector components, and the different $p_T$ distributions for the different processes.

\section*{Usage Notes}

The dataset publication is accompanied by a {\tt Python}-based software package, providing examples on how to read the data, train an AD algorithm with them, apply it to the  {\it blackbox} dataset and publish the result (the unique identifier of the 1000 most anomalous events according to some given AD metric)~\cite{GIT_EXAMPLES}. We aim at challenging the interested data scientists within and outside the LHC community to design L1T-friendly AD algorithms, and to submit their list of 1000 most anomalous events to a specific {\tt GitHub} repository~\cite{GIT_RESULTS}. The final goal is to document these algorithms in a dedicated publication and to preserve the benchmark data for future studies. 

\section*{Code availability}
Data are generated using {\tt PYTHIA 8.240}~\cite{pythia}, setting the collision energy at 13\TeV. Unless otherwise specified, all parameters were fixed to their default values.

We set the beam parameters to produce proton-proton collisions at 13~TeV

\begin{verbatim}
Beams:idA = 2212                   ! first beam, p = 2212, pbar = -2212 \\
Beams:idB = 2212                   ! second beam, p = 2212, pbar = -2212 \\
Beams:eCM = 13000.                 ! CM energy of collision \\
\end{verbatim}

while the rest of the card is configured specifically for each process, as indicated in the {\tt PYTHIA} manual~\cite{pythia}. For example, $W \to \ell \nu$  decays are generating setting:

\begin{verbatim}
WeakSingleBoson:ffbar2W = on         ! switch on W production mode
24::onMode = off                     ! switch off any W decay
24:onIfAny = 11 13 15                ! switch on W-> lv decays.
\end{verbatim}

The signal-specific parameters for the four benchmark signal models are set as follows:
\begin{itemize}
    \item For $A \to 4\ell$: set the Higgs mass to 50 GeV, force the decay to $Z^*Z^*$ final states, and force $Z^* \to \ell \ell$ decays ($\ell=e, \mu, \tau$).
    \item For $LQ \to b \tau$: set the $LQ$ mass to 80~GeV and force its decays to a $b$ quark and a $\tau$ lepton.
    \item For $h^0 \to \tau \tau$: set the Higgs boson mass to 60~GeV and switch off any decay mode other than $\tau \tau$.
    \item For $h^+\to \tau \nu$: set the charged Higgs boson mass to 60~GeV and switch off any decay mode other than $\tau \nu$.
\end{itemize}

We emulate detector response with {\tt DELPHES 3.3.2}~\cite{deFavereau:2013fsa},  using the default Phase-II CMS detector card. For simplicity, we avoid degrading the detector resolution to account for the coarser nature of L1T event reconstruction. This simplification does not affect the aim of the study, which is not focused on assessing the absolute physics performance but instead on comparing different algorithms and their resource consumption. We include the effect of parasitic proton collisions, sampling the number of collisions according to a Poisson distribution centered at 20. The Delphes outcome is processed by a custom {\tt Python} macro to store the aforementioned physics content on HDF5 files, which are then published.

\bibliography{biblio}

\section*{Acknowledgements}

This project has received funding from the European Research Council (ERC) under the European Union's Horizon 2020 research and innovation programme (grant agreement No. 772369) and the ERC-POC programme (grant No. 996696). 

\section*{Author contributions statement}

J.N. conceived the idea of publishing the dataset and creating a data challenge on it; M.P. created the data in raw format; E.P. and E.G. applied the event selection and produced the dataset in its final format; T.A., J.N. and K.W. conceived the package with example code; E.P. designed the example autoencoder; all drafted the paper.

\section*{Competing interests}

The authors declare no competing interests.

\end{document}